\documentclass[runningheads]{llncs}
\usepackage[T1]{fontenc}
\usepackage{multirow}
\usepackage{booktabs}
\usepackage[colorlinks,linkcolor=red]{hyperref}
\usepackage{amsmath}
\usepackage{wrapfig}
\usepackage{lipsum}
\usepackage{amsfonts}
\usepackage{algorithm}
\usepackage[noend]{algpseudocode}
\usepackage{graphicx}
\graphicspath{{./img/}}
\DeclareMathOperator*{\argmin}{arg\,min}

\begin{document}
%
\title{InverseSR: 3D Brain MRI Super-Resolution Using a Latent Diffusion Model}
%
%
\author{Jueqi Wang\inst{1} \and
Jacob Levman\inst{1,2,3} \and
Walter Hugo Lopez Pinaya\inst{4} \and
Petru-Daniel Tudosiu\inst{4} \and
M. Jorge Cardoso\inst{4} \and
Razvan Marinescu\inst{5}}


%
\institute{St. Francis Xavier University, Canada \\
\email{\{x2019cwn,jlevman\}@stfx.ca} \and
Martinos Center for Biomedical Imaging, Department of Radiology, Massachusetts General Hospital, Charlestown, USA \and
Nova Scotia Health Authority \and
School of Biomedical Engineering \& Imaging Sciences, King’s College London, UK \\
\email{\{walter.diaz\_sanz, petru.tudosiu,\\ m.jorge.cardoso\}@kcl.ac.uk} \and
University of California, Santa Cruz, USA \\
\email{ramarine@ucsc.edu}
}

\maketitle  
\begin{abstract}
High-resolution (HR) MRI scans obtained from research-grade medical centers provide precise information about imaged tissues. However, routine clinical MRI scans are typically in low-resolution (LR) and vary greatly in contrast and spatial resolution due to the adjustments of the scanning parameters to the local needs of the medical center. End-to-end deep learning methods for MRI super-resolution (SR) have been proposed, but they require re-training each time there is a shift in the input distribution. To address this issue, we propose a novel approach that leverages a state-of-the-art 3D brain generative model, the latent diffusion model (LDM) from \cite{10.1007/978-3-031-18576-2_12} trained on UK BioBank, to increase the resolution of clinical MRI scans. The LDM acts as a generative prior, which has the ability to capture the prior distribution of 3D T1-weighted brain MRI. Based on the architecture of the brain LDM, we find that different methods are suitable for different settings of MRI SR, and thus propose two novel strategies: 1) for SR with more sparsity, we invert through both the decoder $\mathcal{D}$ of the LDM and also through a deterministic Denoising Diffusion Implicit Models (DDIM), an approach we will call InverseSR(LDM); 2) for SR with less sparsity, we invert only through the LDM decoder $\mathcal{D}$, an approach we will call InverseSR(Decoder). These two approaches search different latent spaces in the LDM model to find the optimal latent code to map the given LR MRI into HR. The training process of the generative model is independent of the MRI under-sampling process, ensuring the generalization of our method to many MRI SR problems with different input measurements. We validate our method on over 100 brain T1w MRIs from the IXI dataset. Our method can demonstrate that powerful priors given by LDM can be used for MRI reconstruction. Our source code is available online: \url{https://github.com/BioMedAI-UCSC/InverseSR}.

\end{abstract}
\section{Introduction}
\indent End-to-end convolutional neural networks (CNNs) have shown remarkable performance compared to classical algorithms \cite{iglesias2021joint} on MRI SR. Deep CNNs have been widely applied in a variety of MRI SR situations; for instance, slice imputation on the brain, liver and prostate MRI \cite{10.1016/j.compbiomed.2022.105667} and brain MRI SR reconstruction on scaling factors $\times 2$, $\times 3$, $\times 4$ \cite{zhang2021mr}. Several techniques based on deep CNNs have been proposed to improve performance, such as densely connected networks \cite{8363679}, adversarial networks \cite{chen2018efficient}, and attention network \cite{zhang2021mr}. However, their supervised training requires paired images, which necessitates re-training every time there is a shift in the input distribution \cite{CHEN2018446,KAMNITSAS201761}. As a result, such methods are unsuitable for MRI SR, as it is challenging to obtain paired training data that cover the variability in acquisition protocols and resolution of clinical brain MRI scans across institutions \cite{iglesias2021joint}.\\
\indent Building image priors through generative models has recently become a popular approach in the field of image SR, for both computer vision \cite{bora2017compressed,abdal2019image2stylegan,menon2020pulse,chung2022diffusion,lugmayr2022repaint} as well as medical imaging \cite{marinescu2020bayesian,song2021solving}, as they do not require re-training in the presence of several types of input distribution shifts. While these methods have shown promise in MRI SR, they have so far been limited to 2D slices \cite{marinescu2020bayesian,song2021solving}, rendering them unsuitable for 3D brain MRIs slice imputation.\\
\indent In this study, we propose solving the MRI SR problem by building powerful, 3D-native image priors through a recently proposed HR image generative model, the latent diffusion model (LDM) \cite{10.1007/978-3-031-18576-2_12,rombach2022high}. We solve the inverse problem by finding the optimal latent code $z$ in the latent space of the pre-trained generative model, which could restore a given LR MRI $I$, using a known corruption function $f$. In this study, we focus on slice imputation, yet our method could be applied to other medical image SR problems by implementing different corruption functions $f$. We proposed two novel strategies for MRI SR: Inverse(LDM), which additionally inverts the input image through the deterministic DDIM model, and InverseSR(Decoder) which inverts the input image through the corruption function $f$ and through the decoder $\mathcal{D}$ of the LDM model. We found that for large sparsity, InverseSR(LDM) had a better performance, while for low sparsity, InverseSR(Decoder) performed best. While the LDM model was trained on UK BioBank, we demonstrate our methods on an external dataset (IXI) which was inaccessible to the pre-trained generative model. Both quantitative and qualitative results show that our method achieves significantly better performance compared to two other baseline models. Furthermore, our method can also be applied to tumour/lesion filling by creating tumour/lesion shape masks.

\subsection{Related work}
\subsubsection{MRI Super-resolution} End-to-end deep training \cite{9093603,10.1016/j.compbiomed.2022.105667,zhang2021mr} has been proposed recently for MRI SR, which has achieved superior results compared to classical methods. However, these methods require paired data to train, which is hard to acquire because of the large variability present in clinical MRIs  \cite{SANDER2022102393,iglesias2021joint}. To circumvent this limitation, several unsupervised methods have been proposed without requiring access to HR scans \cite{dalca2018medical,iglesias2021joint,brudfors2019tool}. Dalca et al. \cite{dalca2018medical} proposed a gaussian mixture model for sparse image patches. Brudfors et al. \cite{brudfors2019tool} presented an algorithm which could take advantage of multimodal MRI. Iglesias et al. \cite{iglesias2021joint} introduced a method to train a CNN for MRI SR on any given combination of contrasts, resolutions and orientations.

\subsubsection{Solving inverse problems using generative models} A common way to solve the inverse problem using an LDM is to use the encoder $\mathcal{E}$ to first encode the given image $x$ into the latent space $z_0=\mathcal{E}(x)$ \cite{elarabawy2022direct,gal2022image,mokady2022null}, followed by DDIM (Denoising Diffusion Implicit Models) Inversion \cite{dhariwal2021diffusion,song2020denoising} to encode $z_0$ into the noise latent code $z_T$ \cite{mokady2022null}. However, this approach does not work for low-resolution images, because the encoder $\mathcal{E}$ has only been trained on high-resolution images.

Our work is also similar to the optimization-based generative adversarial network (GAN) inversion approach \cite{Xia2021GANIA}, trying to find the optimal latent representation $z^*$ in the latent space of GAN, which could be mapped to represent the given image $x \approx G(z^*)$. More recent works \cite{chung2022diffusion,elarabawy2022direct,hertz2022prompt,lugmayr2022repaint,song2021solving} have used diffusion models for inverse problems due to their superior performance. However, all these methods require the diffusion model to operate directly in the image space, which for large image resolutions can become GPU-memory intensive.

\section{Methods}
\subsubsection{3D Brain Latent Diffusion Models} 
\begin{figure}
\includegraphics[width=\textwidth]{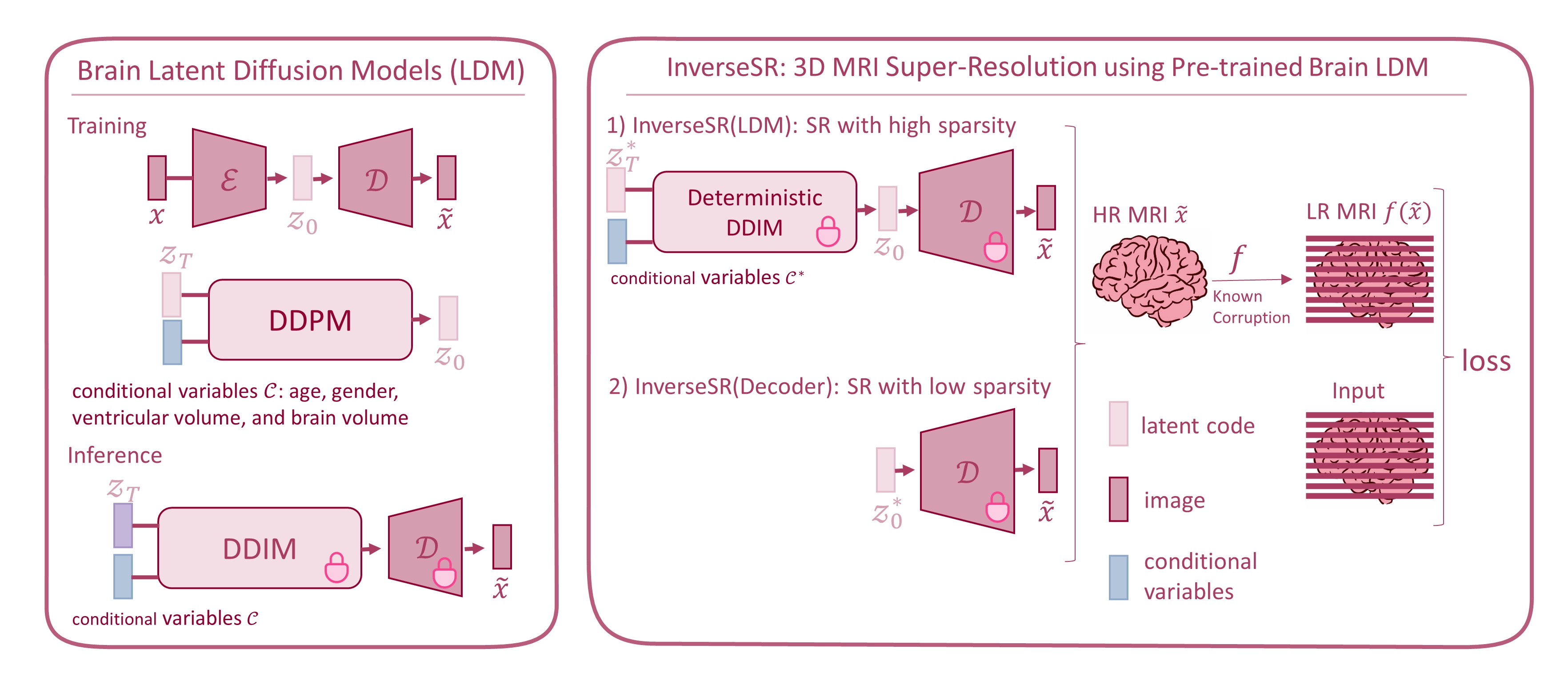}
\caption{(Left) The Brain LDM has two stages of training process. First, an autoencoder is pre-trained to map T1w brain MRIs into a latent code $z_0=\mathcal{E}(x)$. Then, a diffusion model is trained to generate $z_0$ latents from this learned latent space. During inference, DDIM has been applied to reduce the sampling step with little performance drop. (Right) We proposed two methods to handle different scenarios of MRI SR, based on the architecture of brain LDM: 1) InverseSR(LDM): for SR with high sparsity, we optimize the latent code $z_T^*$ and associated conditional variables $\mathcal{C^*}$ using deterministic DDIM and decoder $\mathcal{D}$ to map the latent code into brain MRI. 2) InverseSR(Decoder): for SR with low sparsity, we optimize the $z_0^*$ which only use the decoder $\mathcal{D}$ to map the latent code into brain MRI.} \label{fig1}
\end{figure}

We leverage a state-of-the-art LDM \cite{10.1007/978-3-031-18576-2_12} to create high-quality priors for 3D brain MRIs. There are two components in an LDM: an autoencoder and a diffusion model \cite{rombach2022high}. An encoder $\mathcal{E}$ maps each high-resolution T1w brain MRI $x \sim p_{data}(x)$ into a latent vector $z_0 = \mathcal{E}(x)$ of size $20 \times 28 \times 20$. The decoder $\mathcal{D}$ is trained to map the latent vectors $z_0$ back into the MRI image domain $x$. The autoencoder was trained on 31,740 T1w MRIs from the UK Biobank \cite{10.1371/journal.pmed.1001779} using a combination of an L1 loss, a perceptual loss \cite{zhang2018unreasonable}, a patch-based adversarial loss \cite{esser2021taming} and a KL regularization term in the latent space. The autoencoder was trained on pre-processed MRIs using UniRes \cite{brudfors2019tool} into a common MNI space with a voxel size of $1mm^3$ and was then kept unchanged during the LDM training. The latent representations of the T1w brain MRIs were then used to train the LDM. A conditional U-Net $\epsilon_\theta$ was then trained to predict the artificial noise by the following objective:
\begin{equation}
\theta^*=\argmin_{\theta} \mathbb{E}_{z \sim \mathcal{E}(x),\epsilon \sim \mathcal{N}(0,1),t}||\epsilon - \epsilon_\theta(z_t,t,\mathcal{C})||_2^2
\end{equation}
DDIM \cite{song2020denoising} has been used in brain LDM to replace the denoising diffusion probabilistic models (DDPM) during inference to reduce the number of reverse steps with minimal performance loss \cite{10.1007/978-3-031-18576-2_12,song2020denoising}. This network $\varepsilon_\theta$ is conditioned on four conditional variables $\mathcal{C}$: \textit{age, gender, ventricular volume} and \textit{brain volume}, which are all introduced by cross-attention layers \cite{rombach2022high}. Gender is a binary variable, while the rest of the covariates are scaled to $[0, 1]$. Finally, the pre-trained decoder maps the latent vector into an HR MRI $\tilde{x}=\mathcal{D}(z_0)$. The architecture of the brain LDM can be found in Fig.~\ref{fig1}.

\subsubsection{Deterministic DDIM Sampling} In order to obtain a latent representation $z_T$ capable of reconstructing a given noisy sample into a high-resolution image, we employ deterministic DDIM sampling \cite{song2020denoising}:
\begin{equation}
z_{t-1} = \sqrt{\alpha_{t-1}}\left(\frac{z_t - \sqrt{1-\alpha_t}\cdot\epsilon_\theta(z_t,\mathcal{C},t)}{\sqrt{\alpha_t}}\right) + \sqrt{1-\alpha_{t-1}}\cdot\epsilon_\theta(z_t, \mathcal{C}, t)\label{DDIM}
\end{equation}
where $\alpha_{1:T}\in(0,1]^T$ is a time-dependent decreasing sequence,  $\frac{z_t - \sqrt{1-\alpha_t}\cdot\epsilon_\theta(z_t,\mathcal{C},t)}{\sqrt{\alpha_t}}$ represents the ``predicted $x_0$``, and $\sqrt{1-\alpha_{t-1}}\cdot\epsilon_\theta(z_t, \mathcal{C}, t)$ can be understood as the ``direction pointing to $x_t$`` \cite{song2020denoising}.

\subsubsection{Corruption function $f$} We assume a corruption function $f$ known \emph{a-priori} that is applied on the HR image $\tilde{x}$ obtained from the generative model, and compute the loss function based on the corrupted image $f \circ \tilde{x}$ and the given LR input image $I$. In clinical practice, a prevalent method for acquiring MR images is prioritizing high in-plane resolution while sacrificing through-plane resolution to expedite the acquisition process and reduce motion artifacts \cite{zhao2020smore}. To account for this procedure, we introduce a corruption function that generates masks for non-acquired slices, enabling our method to in-paint the missing slices. For instance, on $1\times1\times4 \text{mm}^3$ undersampled volumes, we create masks for three slices every four slices on the generated HR $1\times1\times1 \text{mm}^3$ volumes.

\begin{algorithm}[H]\caption{InverseSR(LDM)}\label{alg:InverseSR}
\begin{algorithmic}[1]
      \State \textbf{Input}: Low-resolution MR image $I$.
      \State \textbf{Output}: Optimized noise latent code $z_T^*$ and conditional variables $\mathcal{C}^*$\vspace{1mm}
      \vspace{1mm}
      \State Initialize $z_T$ with gaussian noise from $N(0,\mathbb{I})$;
      \State Initialize conditional variables $\mathcal{C}=0.5$;
      \For{$j = 0, \dots, N-1$}
        \For{$t=T, T-1, \dots, 1$}
                \State $z_{t-1} \gets \sqrt{\alpha_{t-1}}\left(\frac{z_t - \sqrt{1-\alpha_t}\cdot\epsilon_\theta(z_t,\mathcal{C},t)}{\sqrt{\alpha_t}}\right) + \sqrt{1-\alpha_{t-1}}\cdot\epsilon_\theta(z_t, \mathcal{C}, t)$;
        \EndFor
        \State $L \gets \lambda_{perc}L_{perc}(f \circ \mathcal{D}(z_0), I) + \lambda_{mae} \|f \circ \mathcal{D}(z_0) - I\|$;
        \State $\mathcal{C} \gets \mathcal{C} - \alpha \nabla_{\mathcal{C}}L$;
        \State $z_T \gets z_T - \alpha \nabla_{z_T}L$;
      \EndFor
      \State Set $z_T^* \gets z_T$; $\mathcal{C}^* \gets \mathcal{C}$;
    \State \textbf{Return} $z_T^*,\mathcal{C}^*$;
\end{algorithmic}
\end{algorithm}

\subsubsection{InverseSR(LDM):} In the case of high sparsity MRI SR, we optimize the noise latent code $z_T^*$ and its associated conditional variables $\mathcal{C^*}$ to restore the HR image from the given LR input image $I$ using the optimization method:
\begin{align}
\begin{split}
z_T^*, \mathcal{C}^*=\argmin_{z_T, \mathcal{C}}&\lambda_{perc}L_{perc}(f \circ \mathcal{D}(\text{DDIM}(z_T, \mathcal{C},T)), I) +\\ & \lambda_{mae} \|f \circ \mathcal{D}(\text{DDIM}(z_T, \mathcal{C},T)) - I\|
\end{split}
\end{align}
where $\text{DDIM}(z_T,\mathcal{C},T)$ represents $T$ deterministic DDIM sampling steps on the latent $z_0$ in Equation~\ref{DDIM}. We follow the brain LDM model to use the perceptual loss $L_{perc}$ and the L1 pixelwise loss. The loss function is computed on the corrupted image generated from the generative model and the given LR input. A detailed pseudocode description of this method can be found in algorithm~\ref{alg:InverseSR}.

\subsubsection{InverseSR(Decoder):} For low sparsity MRI SR, we directly find the optimal latent code $z_T^*$ using the decoder $\mathcal{D}$:
\begin{equation}
z_0^*=\argmin_{z_0}\lambda_{perc}L_{perc}(f \circ \mathcal{D}(z_0), I) + \lambda_{mae} \|f \circ \mathcal{D}(z_0) - I\|
\end{equation}

\section{Experimental Design}

\subsubsection{Dataset For Validation:} We use 100 HR T1 MRIs from the IXI dataset (\url{http://brain-development.org/ixi-dataset/}) to validate our method, after filtering out those scans where registration failed. We note that subjects in the IXI dataset are around 10 years younger on average than those in UK Biobank. The MRI scans from UK Biobank also had the faces masked out, while the scans from IXI did not. This caused the faces of our reconstructions to appear blurred.

\begin{figure}
\includegraphics[width=\textwidth]{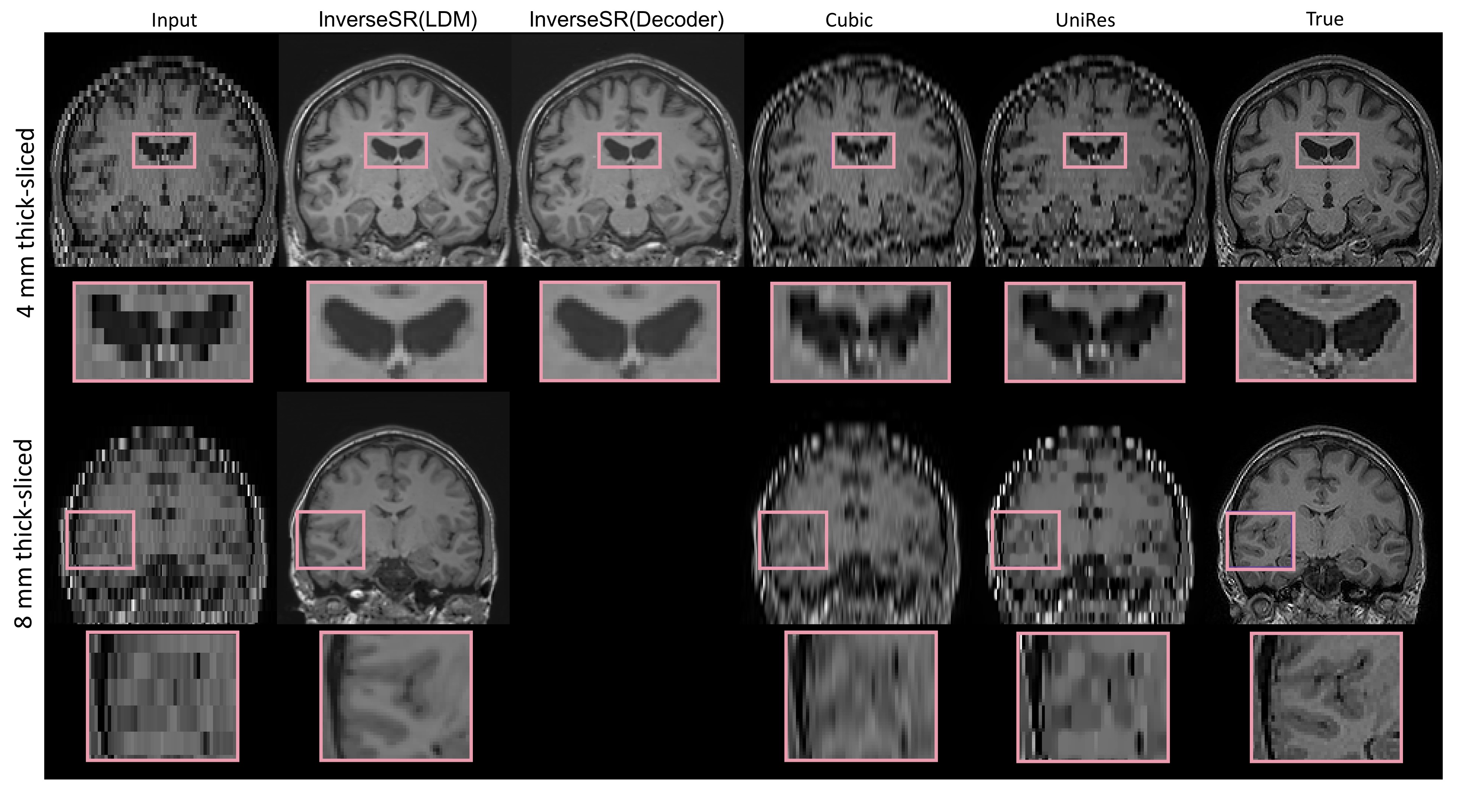}
\caption{Qualitative results of our approach (InverseSR) and the cubic and UniRes baselines on scans with 4 $mm$ and 8 $mm$ thickness.} \label{fig2}
\end{figure}

\subsubsection{Implementation:} Conditional variables are all initialized to 0.5. Voxels in all input volumes are normalized to [0,1]. When sampling the pre-trained brain LDM with the DDIM sampler, we run $T=46$ timesteps due to computational limitations on our hardware. For InverseSR(LDM), $z_T$ is initialized with random gaussian noise. For InverseSR(Decoder), we compute the mean latent code $\bar{z_0}$ as $\bar{z_0}=\sum_{i=1}^{S}\frac{1}{S}\text{DDIM}(z_T^i, \mathcal{C},T)$ by first sampling $S=10,000$ $z_T^i$ samples from $N(0,\mathbb{I})$, then passing them through the DDIM model. $N=600$ gradient descent steps are used for InverseSR(LDM) to guarantee converging (Algorithm \ref{alg:InverseSR}, line 5). 600 optimization steps are also utilized in InverseSR(Decoder). We use the Adam optimizer with $\alpha=0.07$, $\beta_1=0.9$ and $\beta_2=0.999$.

\section{Results}

\indent Fig.~\ref{fig2} shows the qualitative results on the coronal slices of SR from 4 and 8 $mm$ axial scans. The advantage of our approach is clear compared to baseline methods because it is capable of restoring HR MRIs with smoothness even when the slice thickness is large (i.e., 8 $mm$). This is the case because the pre-trained LDM we use is able to build a powerful prior over the HR T1w MRI domain. Therefore, the generated images of our method are HR MRIs with smoothness in 3 directions: axial, sagittal and coronal, no matter how sparse the input images $I$ are. Qualitative results of applying our method on tumour and lesion filling are available in the supplementary material.\\

\indent Table~\ref{tab:quantitativeresult_slice_imputation} shows quantitative results on 100 HR T1 scans from the IXI dataset, which the brain LDM did not have access to during training. We investigated mean peak signal-to-noise ratio (PSNR), and structural similarity index measure (SSIM)~\cite{wang2004image} values and their corresponding standard deviation. We compare our method to cubic interpolation, as well as a similar unsupervised approach,  UniRes \cite{brudfors2019tool}. We show our approach and the two compared methods on two different settings of slice imputation: 4 $mm$ and 8 $mm$ thick-sliced axial scans representing low sparsity and high sparsity LR MRIs, respectively. All the metrics are computed on a 3D volume around the brain of size $160 \times 224 \times 160$. For SR at 4 $mm$, InverseSR(Decoder) achieves the highest mean SSIM and PSNR scores among all compared methods, which are slightly higher than the scores for InverseSR(LDM). For SR at 8 $mm$, Inverse(LDM) achieves the highest mean SSIM and PSNR and lowest standard error than the two baseline methods, which could be attributed to the stronger prior learned by the DDIM model.\\

\begin{table}[b]
\centering
\caption{Quantitative evaluation results (mean $\pm$ standard error) of our approach (InverseSR) and two baselines on 1 mm scans and corresponding SR counterpart - from 4 and 8 $mm$ axial scans.}
\label{tab:quantitativeresult_slice_imputation}
\begin{tabular}{c|c|c|c}
  \toprule
Slice Thickness & Methods & SSIM $\uparrow$ & PSNR $\uparrow$ \\
  \midrule
\multirow{4}{3em}{4 $mm$} & InverseSR(LDM) & 0.797 $\pm$ 0.037 & 28.59 $\pm$ 1.61 \\
& InverseSR(Decoder) & \textbf{0.803 $\pm$ 0.030} & \textbf{29.64 $\pm$ 1.64} \\
& Cubic  & 0.760 $\pm$ 0.052 & 23.84 $\pm$ 2.36 \\
& UniRes~\cite{brudfors2019tool} & 0.688 $\pm$ 0.079 & 21.49 $\pm$ 2.61 \\
\hline
\multirow{4}{3em}{8 $mm$} & InverseSR(LDM) & \textbf{0.754 $\pm$ 0.038} & \textbf{27.92 $\pm$ 1.60} \\
& Cubic  & 0.632 $\pm$ 0.067 & 21.80 $\pm$ 2.36 \\
& UniRes~\cite{brudfors2019tool}   & 0.633 $\pm$ 0.053 & 20.91 $\pm$ 2.29 \\
  \bottomrule
\end{tabular}
\end{table}

\section{Limitations}
One key limitation of our method is the need for large computational resources to perform the image reconstruction, in particular the long Markov chain of sampling steps required by the diffusion model to generate samples. An entire pass through the diffusion model (lines 6-8 in Algorithm \ref{alg:InverseSR}) is required for every step in the gradient descent method. Another limitation of our method is that it is limited by the capacity and output heterogeneity of the LDM generator. 

\section{Conclusions}
In this study, we have developed an unsupervised technique for MRI super-resolution. We leverage a recent pre-trained Brain LDM~\cite{10.1007/978-3-031-18576-2_12} for building powerful image priors over T1w brain MRIs. Unlike end-to-end supervised approaches, which require retraining each time there is a distribution shift over the input, our method is capable of being adapted to different settings of MRI SR problems at test time. This feature is suitable for MRI SR since the acquisition protocols and resolution of clinical brain MRI exams vary across or even within institutions. We proposed two novel strategies for different settings of MRI SR: InverseSR(LDM) for low sparsity MRI and InverseSR(Decoder) for high sparsity MRI. We validated our method on 100 brain T1w MRIs from the IXI dataset through slice imputation using input scans of 4 and 8 $mm$ slice thickness, and compared our method with cubic interpolation and UniRes \cite{brudfors2019tool}.

Experimental results have shown that our approach achieves superior performance compared to the unsupervised baselines, and could create smooth HR images with fine detail even on an external dataset (IXI). Experiments in this paper focus on slice imputation, but our method could be adapted to other MRI under-sampling problems by implementing different corruption functions $f$. For instance, for reconstructing k-space under-sampled MR images, a new corruption function could be designed by first converting the HR image into k-space, then masking a chosen set of k-space measurements, and then converting back to image space. Instead of estimating a single image, future work could also estimate a distribution of reconstructed images through either variational inference (like the BRGM model~\cite{marinescu2020bayesian}) or through sampling methods such as Markov Chain Monte Carlo (MCMC) or Langevin dynamics \cite{jalal2021robust}.

\subsubsection{Acknowledgements} This work was fund by an NSERC Discovery Grant to JL. Funding was also provided by a Nova Scotia Graduate Scholarship and a StFX Graduate Scholarship to JW. Computational resources were provided by Compute Canada.

~\\

%
%
%
\bibliographystyle{splncs04}
\bibliography{mybibliography}
%





\newpage
\begin{Large}
    \begin{center}
            {\bfseries Supplementary Material}
    \end{center}
\end{Large}

\begin{figure}
\includegraphics[width=\textwidth]{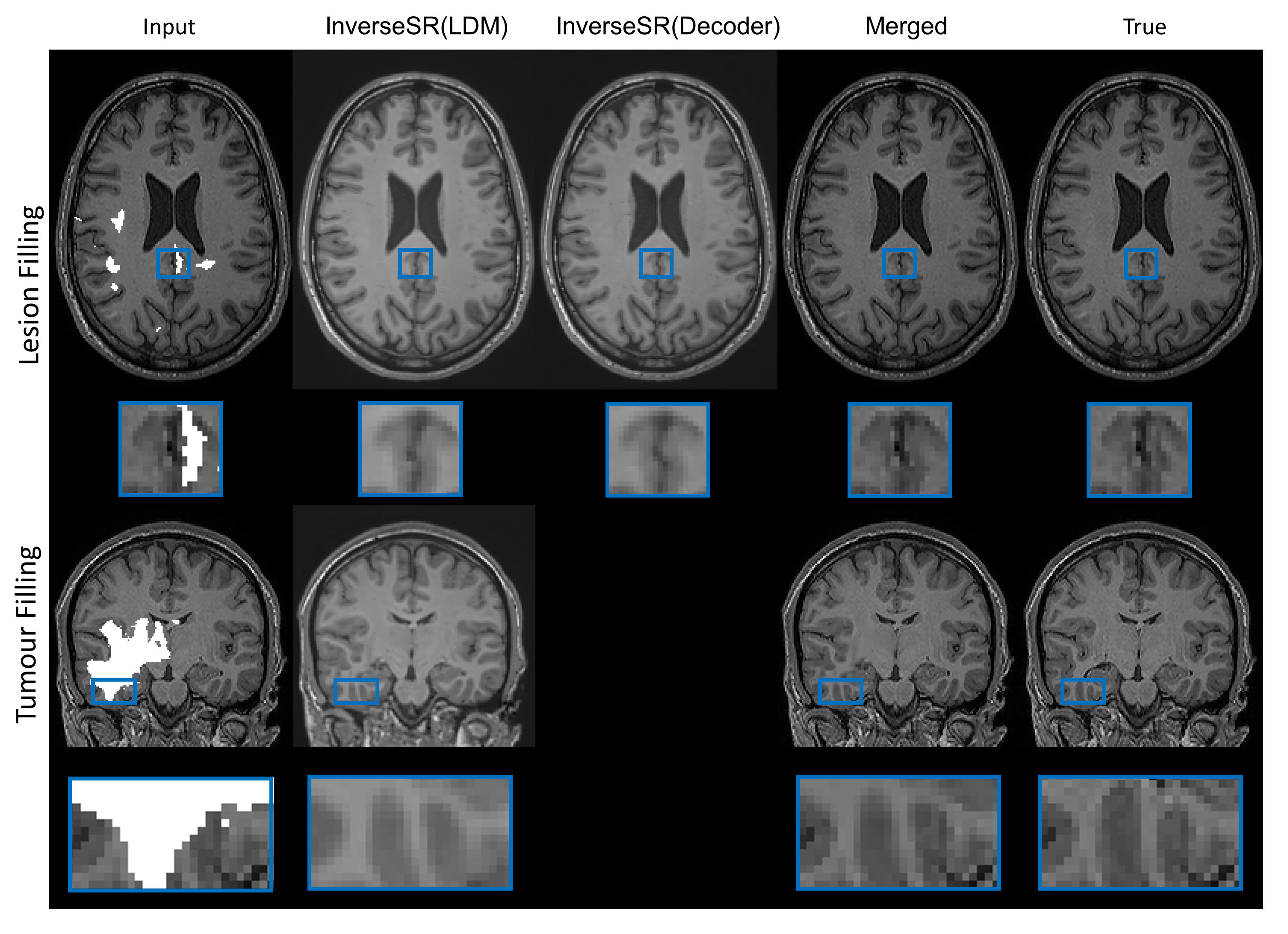}
\caption{Qualitative results of our approach (InverseSR) on lesion and tumour filling. Lesion and tumour masks are separately from the ISBI2015 multiple sclerosis dataset \cite{CARASS201777} and BraTS2020 \cite{6975210}.}
\end{figure}

\end{document}